# Computing recommendations via a Knowledge Graph-aware Autoencoder

Vito Bellini[★], Angelo Schiavone[★], Tommaso Di Noia[★],
Azzurra Ragone[•], Eugenio Di Sciascio[★]
[★]Polytechnic University of Bari
Bari - Italy
firstname.lastname@poliba.it
[•]Independent Researcher
azzurra.ragone@gmail.com

## ABSTRACT

In the last years, deep learning has shown to be a game-changing technology in artificial intelligence thanks to the numerous successes it reached in diverse application fields. Among others, the use of deep learning for the recommendation problem, although new, looks quite promising due to its positive performances in terms of accuracy of recommendation results. In a recommendation setting, in order to predict user ratings on unknown items a possible configuration of a deep neural network is that of autoencoders typically used to produce a lower dimensionality representation of the original data.

In this paper we present KG-AUTOENCODER, an autoencoder that bases the structure of its neural network on the semantics-aware topology of a knowledge graph thus providing a label for neurons in the hidden layer that are eventually used to build a user profile and then compute recommendations. We show the effectiveness of KG-AUTOENCODER in terms of accuracy, diversity and novelty by comparing with state of the art recommendation algorithms.

## 1 INTRODUCTION

Recommender systems (RS) have becoming pervasive tools we experience in our everyday life. While browsing a catalog of items RSs exploit users' past preferences in order to suggest new items they might be interested in. In a digital world where we, as users, are overwhelmed by multiple possibilities and choices they result a valid tool to help us in finding information that fits our need, tastes and preferences. Many online services heavily rely on the usage of a recommender systems to suggest new movies to watch, new books to read or new songs to listen to.

Over the years, different strategies have been proposed to tackle the recommendation problem; among them, collaborative filtering (CF) has shown to be very effective in predicting the relevance of unrated items, especially if many data about users-items interactions are available. CF approaches just use item ratings[1] provided by the users in a system to suggest, in a personalized way, new and unknown items to interact with. Differently from CF RS, content-based (CB) approaches exploit descriptive metadata in order to find items which are similar to the ones already available in a user profile and recommend them accordingly.

Many works [20, 23, 32] show that the recommendation quality can be improved if both strategies are combined in a hybrid one [5]. Exploiting CB methods requires getting information about the items in order to model their corresponding features. In this direction, Knowledge Graphs have been recently widely adopted to represent items, compute their similarity and relatedness [10] as well as to feed CB and hybrid recommendation engines [30]. The publication and spread of freely available Knowledge Graphs in the form of Linked Open Data datasets, such as DBpedia [1], has paved the way to the development of knowledge-aware recommendation engines in many application domains and, still, gives the possibility to easily switch from a domain to another one by just feeding the system with a different subset of the original graph.

Another technology that surely boosted the development of a new generation of smarter and more accurate recommender systems is deep learning [6]. Starting from the basic notion of artificial neural net (ANN), several deep learning strategies have been developed to address classical problems in artificial intelligence, like image recognition or natural language processing. Among the different configuration of a deep ANN, here we are interested in autoencoders. Initially conceived for feature selection and dimensionality reduction [44] they have been then used as generative models of data [21].

In this paper we show how autoencoders technology can benefit from the existence of a Knowledge Graph to create a representation of a user profile that can be eventually exploited to predict ratings for unknown items. The main intuition behind the approach is that both ANN and Knowledge Graph expose a graph-based structure. Hence, we may imagine to build the topology of the inner layers in the ANN by mimicking that of a Knowledge Graph. We will see how this idea can lead to an ANN whose nodes in the hidden layers have an explicit label and semantics attached that can be further exploited to represent user preferences. In fact, the topolgy of such a neural network is built upon the semantic interconnections between items and categories which exist in the adopted knowledge graph. Once trained, the values obtained for each hidden neuron of the ANN are interpreted as the relevance that the associated feature has for the user.

---

[1]Here, with ratings we refer to whatever user interaction, both implicit and explicit, from which we can infer a like or dislike behavior.

The remainder of this paper is structured as follows: in the next section we discuss related works on recommender systems exploiting deep learning, knowledge graphs and Linked Open Data. Then, the basic notions of the technologies we adopted are introduced in Section 3. The proposed recommendation model is described in Section 4 while in Section 5 we present the experimental setting and evaluation. Conclusions and Future Work close the paper.

## 2 RELATED WORK

**Autoencoders and Deep Learning for RS.** The adoption of deep learning techniques is for sure one the main advances of the last years in the field of recommender systems. In [47], the authors propose the usage of a denosing autoencoder to perform a top-N recommendation task. Denoising autoencoders, exploiting a corrupted version of the input data, learn the latent knowledge behind it; in the same way, the authors corrupt data about rated items in order to improve users' preferences thus learning and finding the most interesting items for them. A collaborative filtering model based on autoencoders is described in [39], in which the authors develop both user-based and item-based autoencoders to tackle the recommendation task. Stacked Denoising Autoencoders are combined with collaborative filtering techniques in [42] where the authors leverage autoencoders to get a smaller and non-linear representation of the users-items interactions. This representation is eventually used to feed a deep neural network which can alleviate the cold-stat problem thanks to the integration of side information. A hybrid recommender system is finally built. Another hybrid approach is proposed in [12]: the aim is to address the sparsity problem using deep learning techniques to model both users and items exploiting side information. The representations so obtained are then integrated in a collaborative model relying on matrix factorization. Wang et al. [46] suggest to apply deep learning methods on side information to reduce the sparsity of the ratings matrix in collaborative approaches. In [45] the authors propose a deep learning approach to build a high-dimensional semantic space based on the substitutability of items; then, a user-specific transformation is learnt in order to get a ranking of items from such a space. Analysis about the impact of deep learning on both recommendation quality and system scalability are presented in [13], where the authors first represent users and items through a rich feature set made on different domains and then map them to a latent space. Finally, a content-based recommender system is built.

**Knowledge Graphs and Linked Open Data for RS.** Several works have been proposed exploiting side information coming from knowledge graphs and Linked Open Data (LOD) to enhance the performance of recommender systems. Most of them rely on the usage of DBpedia as knowledge graph. In [18], for the very first time, a LOD-based recommender system is proposed to alleviate some of the major problems that affect collaborative techniques mainly the high sparsity of the user-item matrix. The effectiveness of such an approach seems to be confirmed by the large number of methods that have been proposed afterwards. A detailed review of LOD-based recommender systems is presented in [7]. By leveraging the knowledge encoded in DBpedia, it is possible to build an accurate content-based recommender system [9] also in mobile scenarios [33] or a multirelational graph for graph-based recommenders [11]. Even cross-domain recommendations [15] may be easily provided, letting users get suggestions for items belonging to different domains, such as movies and songs. Furthermore, the exploitation of Linked Open Data helps to deal with *limited content analysis* and *cold-start* problems: new relevant features can be introduced to improve item representations [4, 38] or to cope with the increasing data sparsity [27, 43].

It is worth noticing that leveraging knowledge graphs available as Linked Open Data is also useful to improve the overall accuracy of a recommender [26, 32] or to provide a good balance between different recommendation objectives, such as accuracy, diversity and novelty [20, 27, 31].

Further applications of DBpedia properties embrace some interesting tasks like the generation of effective natural-language recommendation explanations [28] and the definition of semantic similarity measures for providing more accurate recommendations [23, 29, 35]. Some of the ideas presented in this work have been originally presented in [2] where an analogous approach has been adopted for the first time to tackle the cold start problem.

Vector Space Model (VSM) [37] is an established technique widely used in the fields of information retrieval and information filtering. Over the years, different approaches leveraging on the VSM have been proposed in order to help users in their search for interesting items. Recent works include novel strategies to enrich the user profile modelled by VSM injecting knowledge coming from ontological and semantics-aware data sources: in [14], the authors analyze the impact of ontology and text mining techniques in such a task, leveraging on the synonyms and the hypernyms of the terms to compute each term's weight as a linear combination of three different kinds of TF-IDF values; the enhanced VSM [25] tries to overcome a classical problem of the original VSM: it builds each user profile only considering the associated positive preferences. Therefore, the use of a negation operator is proposed, so that also negative preferences can also be taken into account.

## 3 BACKGROUND TECHNOLOGIES

We now introduce and discuss the main technologies we have been using to model our Knowledge Graph-aware autoencoder. We start by describing the ideas behind autoencoders and their usage for rating prediction in recommendation scenarios and then we give a brief overview on knowledge graphs encoded as LOD with an emphasis on DBpedia.

### 3.1 Autoencoders

An artificial neural network is a mathematical model used to learn the relationships which underlie in a given set of data. Starting from them, after a training phase, an ANN can be used to predict a single value or a vector, for regression or classification tasks.

Basically, an ANN consists of a bunch of nodes, called neurons, distributed among three different kinds of layers: the input layer, one or more hidden layers and the output layer. Typically, a neuron of a layer is connected to all the neurons of the next layer, making the ANN a fully connected network. In Figure 1 we see a graphical representation with an input vector **x** and an output vector **y**. In this case we have only one single hidden layer but an ANN can have, in principle, whatever number of hidden layers.

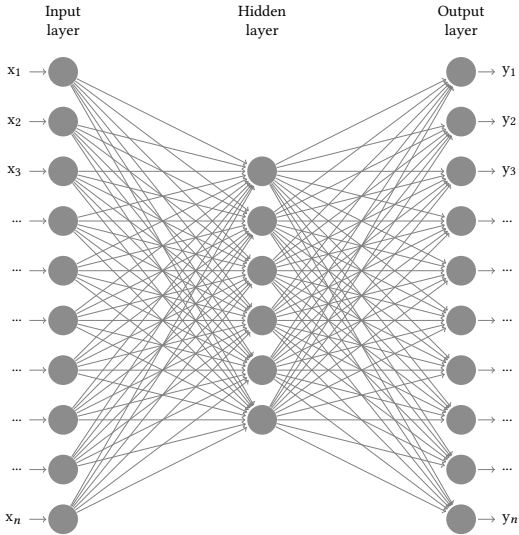

Figure 1: A representation of a fully connected ANN.

Autoencoders are ANNs which try to set the ouput values equal to the input ones, modelling an approximation of the identity function $y = f(x) = x$. Roughly, they are forced to predict the same values they are fed with. Therefore, the number of the output units and the one of the input nodes is the same, i.e. $|x| = |y|$. The aim of such a task is to obtain a new representation of the original data based on the values of the hidden layers neurons. In fact, each of these layers projects the input data in a new Euclidean space whose dimensions depend on the number of the nodes in the hidden layer. Autoencoders have an odd number $h$ of hidden layers. The first half is used to encode the input in the new feature space represented by the middle layer while the second half is used to reconstruct the encoded input in the original feature space. Generally, autoencoders are exploited to get a smaller representation of the input data, using the hidden layers as a bottleneck: it means that the number of neurons decreases from the input data up to the middle layer, in order to encode somehow the original data. Then, the values obtained at the middle layer are used to reconstruct the input data through a decoding operation which involves the next hidden layers: the number of their nodes increases until the input dimensions are reached. As a matter of fact, when we use an autoencoder, we are not interested at its ouput, but at the encoded representation it computes: in this way, we can leverage the implicit knowledge behind the original data, performing the so called feature extraction task. The actual meaning of each dimension (represented by hidden nodes) in the new space is unknown, but we can be sure that they are based on latent patterns binding the training cases. Therefore, autoencoders are mainly used for the purpose of dimensionality reduction, resulting more efficient than other techniques like PCA (Principal Component Analysis) in such cases where non-linear input has to be processed.

*Autoencoders for rating prediction.* As shown in [22], autoencoders can be sucessfully used in collaborative-filtering rating prediction. If both the input and output layer represent the items in a catalog, we may feed the autoencoder with ratings provided by users, in order to let the network learn the latent relations behind them. Therefore, hidden layers will encode a representation of the input data which is based on the ratings provided by all the users in a system. Since both input and output units model all the items in the training set, the autoencoder will be able to compute scores for users' unrated items. Hence, it is possible to provide a top-N recommendation list for each user by feeding the network with her ratings once the autoencoder is trained.

### 3.2 KG

In 2012, Google announced its Knowledge Graph[2] as a new tool to improve the identification and retrieval of entities in return to a search query. Most of the knowledge encoded in Google Knowledge Graph actually came from Freebase. This was a crowdsourced effort to create a base of facts in all possible knowledge domains. A Knowledge Graph is a form of representation of knowledge through a semantic (labelled) network that allows a system to store the human knowledge in a structured format well understandable by a computer agent.

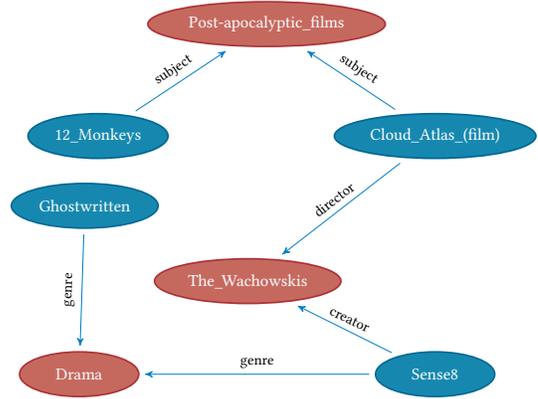

Figure 2: Part of knowledge-graph.

Alongside with the development of the above mentioned initiatives, inspired by the seminal paper by Tim Berners-Lee et al. [3], in the last decade, a set of technologies for the creation of the so called Semantic Web were developed. These technologies are the cornerstone in the development of the Linked Data initiative[3]: an effort to create, interconnect and publish semantic datasets. Among them, the most important is for sure DBpedia[4]. This encodes an important amount of the information available in Wikipedia as RDF triples and is freely available on the Web. If we think that each Wikipedia page corresponds to a unique DBpedia entity, it is easy to see how rich the knowledge available in the DBpedia graph is.

---

[2]https://googleblog.blogspot.it/2012/05/introducing-knowledge-graph-things-not.html
[3]http://linkeddata.org
[4]http://dbpedia.org

Exploiting graph data sources we may investigate relationships among entities and hence discover meaningful paths within the graph.

In figure 2 we show an excerpt of the DBpedia graph, involving some entities in the movie domain. Interestingly we see that DBpedia encodes both factual information, e.g. "*Cloud_Atlas_(film) has director The_Wachowskis*", and categorical one such as "*Cloud_Atlas_(film) has subject Post-apocalyptic_films*".

## 4 SEMANTICS-AWARE AUTOENCODERS FOR RATING PREDICTION

As we have seen in the previous section, autoencoders are unsupervised artificial neural networks able to efficiently reconstruct input data by compressing them in the hidden layers with a low dimensionality representation. Unfortunately, just like other methods for latent representation, they are unable to provide a meaning to the latent factors they provide which are represented by the neurons in the hidden layer. To address this issue, we propose to give an explicit semantics to connections of the neurons within the hidden layer by exploiting information explicitly available in knowledge graphs. The main idea of our approach is therefore to map connections between units from layer i to layer i+1, by mimicking the connections in a knowledge graph (KG) as shown in Figure 3. There we see that we injected only categorical information in the autoencoder and we left out factual one. As a matter of fact, if we analyze these two kinds of information in DBpedia we may notice that:

- the quantity of categorical information is higher than factual one. If we consider movies, the overall number of entities they are related with is lower than the overall number of categories;
- categorical information is more distributed over the items than factual one. Going back to movies we see that they are more connected with each other via categories than via other entities.

Hence, we may argue that for a recommendation task where we are looking for commonalities among items, categorical data may result more meaningful than factual one. The main assumption behind this choice is that, for instance, if a user rated positively *Cloud_Atlas* this may be interpreted as a positive rating for the connected category *Post-apocalyptic_films*.

In order to test our assumption, we mapped the autoencoder network topology with the categorical information related to items rated by users. As we build a different autoencoder for each user depending on the items she rated in the past, the mapping with a Knowledge Graph makes the hidden layer of variable length in the number of units, depending on how much categorical information is available for items rated by the specific user.

Let $n$ be the number of items rated by $u$ available in the graph and $C_i = \{c_{i1}, c_{i2}, \ldots, c_{im}\}$ be the set of $m$ categorical nodes associated in the KG to the item $i$. Then, $F^u = \bigcup_{i=1}^{n} C_i$ is the set of features mapped into the hidden layer for the user and the overall number of hidden units is equal to $|F^u|$. Once the neural network setup is done, the training process takes place, feeding the neural network with ratings provided by the user, normalized in the interval [0,1]. It is worth noticing that, as the autoencoder we build mimic the structure of the connections available in the Knowledge Graph, the neural network we build is not fully connected. Moreover, it does not need bias nodes because these latter are not representative of any semantic data in the graph.

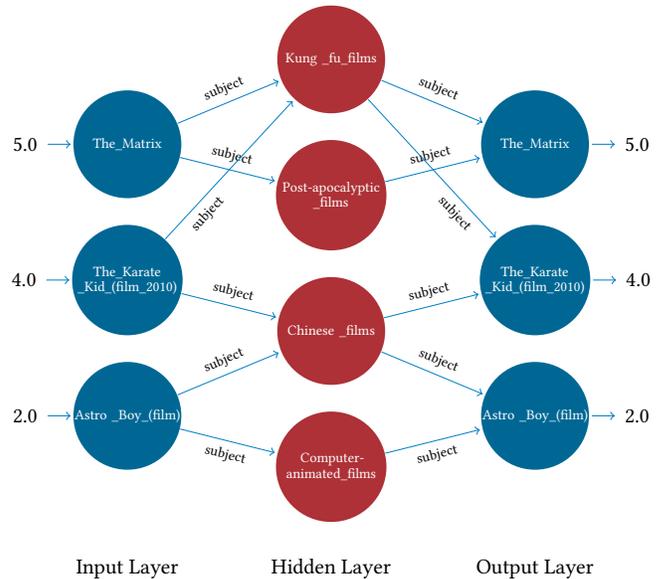

Figure 3: Architecture of a semantic autoencoder.

Nodes in hidden layer correspond to categorical information in the knowledge graph. At every iteration of the training process, backpropagation will change weights accordingly on edges among units in the layers, such that the sum of entering edges in an output unit will reconstruct the user rating for the item represented by that unit. Regarding the nodes in the hidden layer, we may interpret the sum of the weights associated to entering edges computed at the end of the training process as the importance of that feature in the generation of an output which, in our case, are the ratings provided by the user.

### 4.1 User Profiles

Once the network converges we have a latent representation of features associated to a user profile together with their weights. Having one model per user allows us to find the best values of parameters that best reconstruct the user ratings on the output. However, very interestingly, this time the features represented by nodes in the hidden layer also have an explicit meaning as they are in a one to one mapping with categories in a knowledge graph. Our autoencoder is therefore able of learning the semantics behind the ratings of each user and weight them through backpropagation. The structure of a generic hidden unit looks like the one depicted in Figure 4. In our current implementation we used the well known sigmoid $\sigma(x) = \frac{1}{1+e^{-x}}$ activation function because we normalized the design matrix to be within [0, 1] and so we scaled down all the user ratings in that range. We trained each autoencoder for 10,000 epochs with a learning rate of $r = 0.03$; weights are initialized to zero close values as Xavier et al. suggest in [17].

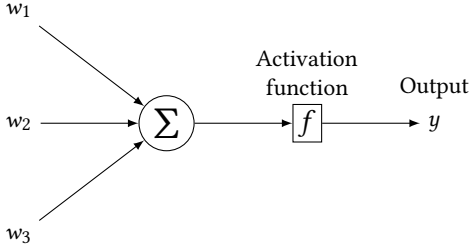

Figure 4: Structure of units

Starting from the trained autoencoder, we may build a user profile by considering the categories associated to the items she rated in the past as features and by assigning them a value according to the weights associated to the edges entering the corresponding hidden units. Given a user $u$, the weight associated to a feature $c$ is then the summation of the weights $w_k^u(c)$ associated to the edges entering the hidden node representing the Knowledge Graph category $c$ after training the autoencoder with the ratings of $u$.

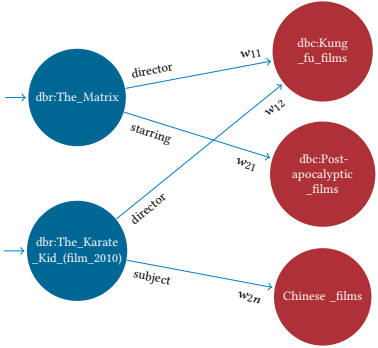

Figure 5: An excerpt of the network in Figure 3 after the training.

More formally, we have:
$$\omega^u(c) = \sum_{k=1}^{|In(c)|} w_k^u(c)$$

where $In(c)$ is the set of the edges entering the node representing the feature $c$. We remember that since the autoencoder is not fully connected, $|In(c)|$ varies depending on the related connections to the category $c$ in the knowledge graph. As an example, if we consider the excerpt of the network in Figure 3 represented in Figure 5, for $c =$ dbc : Kung_fu_films we have:

$$\omega^u(\text{dbc : Kung\_fu\_films}) = w_{11} + w_{12}$$

By means of the weights associated to each feature, we can now model a user profile composed by a vector of weighted categorical features. Given $F^u$ as the set of categories belonging to all the items rated by $u$ and $F = \bigcup_{u \in U} F^u$ as the set of all features among all the users in the system we have for each user $u \in U$ and for each feature $c \in F$:

$$P(u) = \{\langle c, \omega \rangle \mid \omega = \omega^u(c) \text{ if } c \in F^u\}$$

Considering that users provide a different number of ratings, we have an unbalanced distribution in the dimension of user profiles. Moreover, as a user normally rate only a small subset of the entire catalog, we have a huge number of missing features belonging to items not rated by $u$. In order to compute values associated to missing features, we leverage an unsupervised deep learning model inspired by the *word2vec* approach [24]. It is an efficient technique originally conceived to compute word embeddings (i.e. numerical representations of words) by capturing the semantic distribution of textual words in a latent space starting from their distribution within the sentences composing the original text. Given a corpus, e.g. an excerpt from a book, it projects each word in a multidimensional space such that words similar from a semantic point of view result closer to each other. In this way, we are able to evaluate the semantic similarity between two words even if they never appear in the same sentence. Given a sequence of words $[x_1, \ldots, x_n]$ within a window, *word2vec* compute the probability for a new word $x'$ to be next one in the sequence. More formally, it computes $p(x' \mid [x_1, \ldots, x_n])$.

In our scenario, we may imagine to replace sentences represented by sequences of words with user profiles represented by sequences of categories in $c \in F^u$ and then use the *word2vec* approach to compute for a given user $u$ the weight of missing features $c' \notin F^u$.

We need to prepare the user profiles $P(u)$ to be processed by *word2vec*. Hence, we first generate a corpus made of sequences of ordered features where the order is given by $\omega$. The very preliminary step, is that of selecting an order among elements $c \in F^u$ which results coherent for all $u \in U$ thus moving from the set $P(u)$ to a representative sequence of elements $s(u)$.

For each $\langle c, \omega \rangle \in P(u)$ we create a corresponding pair $\langle c, norm(\omega) \rangle$ with *norm* being the mapping function

$$norm : [0, 1] \mapsto \{0.1, 0.2, 0.3, \ldots, 1\}$$

that linearly maps[5] a value in the interval $[0, 1]$ to a real value in the set $\{0.1, 0.2, 0.3, \ldots, 1\}$. The new pairs form the set

$$P^{norm}(u) = \{\langle c, norm(\omega) \rangle \mid \langle c, \omega \rangle \in P(u)\}$$

For each normalized user profile set $P^{norm}(u)$ we then build the corresponding sequence

$$s(u) = [\ldots, \langle c_i, norm(\omega_i^u) \rangle, \ldots \langle c_j, norm(\omega_j^u) \rangle, \ldots]$$

with $\omega_i^u \geq \omega_j^u$.

Once we have the set $S = \{s(u) \mid u \in U\}$ we can feed the *word2vec* algorithm with this corpus in order to find patterns of features according to their distribution across all users. In the prediction phase, by using each user's sequence of features $s(u)$ as input for the trained *word2vec* model, we estimate the probability of $\langle c', norm(\omega') \rangle \in \bigcup_{v \in U} P^{norm}(v) - P^{norm}(u)$ to belong to the given context, or rather to be relevant for $u$. In other words, we compute $p(\langle c', norm(\omega') \rangle \mid s(u))$.

It is worth noticing that given $c' \in F^u$ we may have multiple pairs with $c'$ as first element in $\bigcup_{v \in U} P^{norm}(v) - P^{norm}(u)$. For instance, given the category dbc:Kung_fu_films we may have both $\langle$dbc : Kung_fu_films, 0.2$\rangle$ and $\langle$dbc : Kung_fu_films, 0.5$\rangle$, with the corresponding probabilities $p(\langle$dbc : Kung_fu_films, 0.2$\rangle \mid s(u))$, $p(\langle$dbc : Kung_fu_films, 0.5$\rangle \mid s(u))$. Still, as we want to add the category dbc:Kung_fu_films together with its corresponding weight only once in the user profile we select only the pair with the highest probability. The new user profile is then

$$\hat{P}(u) = P(u) \cup \{\langle c, \omega \rangle \mid \underset{\omega \in \{0.1, \ldots, 1\}}{\mathrm{argmax}}\ p(\langle c, \omega \rangle \mid s(u))\ \text{and}\ \langle c, \omega \rangle \notin P^{norm}(u)\}$$

---
[5]In our current implementation we use a standard minmax normalization.

We point out that while the original $P(u)$ is built by exploiting only content-based information, the enhanced user profile $\hat{P}(u)$ also considers collaborative information as it based also on the set $S$ containing a representation for the profiles of all the users in $U$.

## 4.2 Computing Recommendations

Given the user profiles represented as vectors of weighted features, recommendations are then computed by using a well-known k-nearest neighbors approach. User similarities are found through projecting their user profile in a Vector Space Model, and then similarities between each pair of users $u$ and $v$ is computed using the cosine similarity:

$$sim(u, v) = \frac{P(u) \cdot P(v)}{||P(u)|| \cdot ||P(v)||} \quad (1)$$

For each user $u$ we find the top-k similar neighbors to infer the rate $r$ for the item $i$ as the weighted average rate that the neighborhood gave to it:

$$r(u, i) = \frac{\sum_{j=1}^{k} sim(u, v_j) \cdot r(v_j, i)}{\sum_{j=1}^{k} sim(u, v_j)} \quad (2)$$

where $r(v_j, i)$ is the rating assigned to $i$ by the user $v_j$. We use then ratings from Equation (2) to provide top-N recommendation for each user.

## 5 EXPERIMENTS

In this section, we present the experimental evaluations performed on three different datasets. We first describe the structure of the datasets used in the experiments and the evaluation protocol and then we move to the metrics adopted for the evaluation and the discussion of obtained results.

Our experiments can be reproduced through the implementation available on our public repository[6].

### 5.1 Dataset

In order to validate our approach we performed experiments on the three datasets summarized in Table 1.

|  | #users | #items | #ratings | sparsity |
| --- | --- | --- | --- | --- |
| MovieLens 20M | 138,493 | 26,744 | 20,000,263 | 99.46% |
| Amazon Digital Music | 478,235 | 266,414 | 836,006 | 99.99% |
| LibraryThing | 7,279 | 37,232 | 626,000 | 99.77% |

Table 1: Datasets

In MovieLens 20M dataset, each user has at least 20 ratings which are made on a 5-star scale; in Amazon Digital Music users express their interest in musical albums with a value in 1-5 range as well; in LibraryThing rates are made on a 10-star scale.

In our experiments, we referred to the freely available knowledge graph of DBpedia[7]. The mapping contains 22,959 mapped items for MovieLens 20M, 4,077 items mapped for Amazon Digital Music and 9,926 items mapped for LibraryThing. For our experiments, we removed from the datasets all the items without a mapping in DBpedia.

### 5.2 Evaluation protocol

Here, we show how we evaluated performances of our methods in recommending items. We split the dataset using Hold-Out 80/20, ensuring that every user has 80% of their ratings in the training set and the remaining 20% in the test set. For the evaluation of our approach we adopted the "all unrated items" protocol described in [41]: for each user $u$, a top-N recommendation list is provided by computing a score for every item $i$ not rated by $u$, whether $i$ appears in the user test set or not. Then, recommendation lists are compared with the test set by computing performance metrics.

### 5.3 Metrics

In this work we avoided to use Root Mean Squared Error (RMSE) because a good predictor is not necessarily a good recommender. It is known that it may estimate the same error for top-N items and bottom-N items, without taking into account that an error in top-N items should be more relevant compared to an error for lower ranked items. For this reason, we chose to use Precision, Recall and nDCG to evaluate the accuracy of our model in recommendation scenarios. To be more precise, we focused the evaluation on F1-Score, because it combines both Precison and Recall in a harmonic mean.

Precision is defined as the fraction of retrieved items that are relevant to the user.

$$Precision@N = \frac{|L_u(N) \cap TS_u^+|}{N}$$

where $L_u(N)$ is the recommendation list up to the N-th element and $TS_u^+$ is the set of relevant test items for $u$. Precision measures the system's ability to reject any non-relevant documents in the retrieved set.

Recall is defined as the fraction of relevant items that are retrieved.

$$Recall@N = \frac{|L_u(N) \cap TS_u^+|}{TS_u^+}$$

Recall measures the system's ability to find all the relevant documents.

Recall increases when a relevant document is retrieved but this causes Precision to decrease, for that reason is necessary to find out a good trade-off between the two measures. Precision and recall can be therefore combined with each other in the F1 measure computed as the harmonic mean between precision and recall.

$$F1@N = 2 \cdot \frac{Precision@N \cdot Recall@N}{Precision@N + Recall@N}$$

---

[6]https://github.com/sisinflab/SEMAUTO-2.0

[7]https://dbpedia.org

In information retrieval, Discounted cumulative gain (DCG) is a metric of ranking quality that measures the usefulness of a document based on its position in the result list. It is based on two assumptions: the former is that high relevant documents are more useful than marginally relevant documents, the latter is that lower ranked relevant documents means less usefulness for user since it is less likely to be consumed. Recommended results may vary in length depending on the user, therefore is not possibile to compare performances among different users, so the cumulative gain at each position should be normalized across users. Hence, normalized discounted cumulative gain, or nDCG, is computed as:

$$nDCG_u@N = \frac{1}{IDCG@N} \sum_{p=1}^{N} \frac{2^{r_{up}} - 1}{\log_2(1 + p)}$$

where $p$ is the position of an item in the recommendation list and $IDCG@N$ indicates the score obtained by an ideal ranking of $L_u(N)$.

Accuracy metrics are a valuable way to evaluate the performance of a recommender system. Nonetheless, it has been argued [40] that also diversity should be taken into account when evaluating users' satisfaction. In order to provide diversified recommendation lists, unpopular items as well as popular ones should be recommended. Aggregate diversity measures how much of the catalog is being consumed:

$$aggrdiv = |\bigcup_{u \in U} L_u(N)|$$

where $L_u(N)$ is the recommendation list up to the N-th element of user $u \in U$. This metric is quite simple because it just counts the unique number of items across all recommendation lists without taking into account how they are distributed among them. Therefore, even if the aggregate diversity metric results to be maximized, some items may be suggested a few times. In order to measure the distribution of items across recommendation lists, Gini index should be used:

$$Gini = \frac{1}{n-1} \sum_{j=1}^{n} (2j - n - 1)p(i_j)$$

where $p(i)$ is the proportion of user choices for item $i$ and $i_1, ...i_n$ is the list of items ordered according to increasing $p(i)$. A Gini index value equal to 0 means that all items are choosen equally often, while it is 1 if a single item is always chosen.

### 5.4 Results Discussion

In our experiments, we compared our approach with three different state of the art techniques widely used in recommendation scenarios: BPRMF, WRMF and a single-layer autoencoder for rating prediction. BPRMF [36] is a Matrix Factorization algorithm which leverages Bayesian Personalized Ranking as objective function. WRMF [19, 34] is a Weighted Regularized Matrix Factorization method which exploits users' implicit feedbacks to provide recommendations. In their basic version, both strategies rely exclusively on the User-Item matrix in a pure collaborative filtering approach. They can be hybridized by exploiting side information, i.e. additional data associated with items. In our experiments, we adopted categorical information found on the DBpedia Knowledge Graph as side information. We used the implementations of BPRMF and WRMF available in MyMediaLite[8] [16] and implemented the autoencoder in Keras[9]. We verified the statistical significance of our experiments by using Wilcoxon Signed Rank test, in fact we get a *p-value* very close to zero, which ensures the validity of our results. In Table 3 we report the results gathered on the three datasets by applying the methods discussed above. As for our approach KG-AUTOENCODER, we tested it for different number of neighbors by varying $k$.

In terms of accuracy, we can see that KG-AUTOENCODER outperforms our baselines on both MovieLens 20M and Amazon Digital Music datasets, while on LibraryThing the achieved results are quite the same. In particular, on the LibraryThing dataset, only the fully-connected autoencoder performs better than our approach with regard to accuracy.

Concerning diversity, we get much better results on all the datasets. Furthermore, by analyzing the gathered results, it seems that our approach provides very discriminative descriptions for each user, letting us to identify the most effective neighborhood and compute both accurate and diversified recommendations. As a matter of fact, we achieve the same results in terms of accuracy as the baselines by suggesting much more items.

As shown in Table 2, KG-AUTOENCODER performs better on those datasets whose items can be associated to a large amount of categorical information, which implies the usage of many hidden units. This occurs because very complex functions can be modeled by ANNs if enough hidden units are provided, as Universal Approximation Theorem points out. For this reason, our approach turned out to work better on MovieLens 20M dataset (whose related neural networks have a high number of hidden units) rather than the others. In particular, the experiments on LibraryThing dataset show that the performances get worse as the number of the neurons decreases, i.e. available categories are not enough.

|  | avg #features | std | avg #features/avg #items |
|---|---|---|---|
| Movielens 20M | 1015.87 | 823.26 | 8.82 |
| Amazon Digital Music | 7.22 | 9.77 | 5.17 |
| LibraryThing | 206.88 | 196.64 | 1.96 |

**Table 2: Summary of hidden units for mapped items only.**

## 6 CONCLUSION AND FUTURE WORK

In this paper, we have presented a recommendation approach that combines the computational power of deep learning with the representational expressiveness of knowledge graphs. As for classical applications of autoencoders to feature selection, we compute a latent representation of items but, in our case, we attach an explicit semantics to selected features. This allows our system to exploit both the power of deep learning techniques and, at the same time, to have a meaningful and understandable representation of the trained model. We used our approach to autoencode user ratings in a recommendation scenario via the DBpedia knowledge graph and proposed an algorithm to compute user profiles then adopted to provide recommendations based on the semantic features we extract with our autoencoder. Experimental results show that we are able to outperform state of the art recommendation algorithms in terms of accuracy and diversity. The results presented in this paper pave the way to various further investigations in different directions. From a methodological and algorithmic point of view, we can surely investigate the augmentation of further deep learning techniques via the injection of explicit and structured knowledge coming from external sources of information. Giving an explicit meaning to neurons in an ANN as well as to their connections can fill the semantic gap in describing models trained via deep learning algorithms. Moreover, having an explicit representation of latent features opens the door to a better and explicit user modeling. We are currently investigating how to exploit the structure of a Knowledge Graph-enabled autoencoder to infer qualitative preferences represented by means of expressive languages such as CP-theories [8]. Providing such a powerful representation may also result in being a key factor in the automatic generation of explanation to recommendation results.

---
[8] http://mymedialite.net
[9] https://keras.io

|                 | k   | F1      | Prec.   | Recall  | nDCG    | Gini    | aggrdiv |
|-----------------|-----|---------|---------|---------|---------|---------|---------|
| **MOVIELENS 20M** ||||||||
| **AUTOENCODER** | –   | 0.21306 | 0.21764 | 0.20868 | 0.24950 | 0.01443 | 1587    |
| **BPRMF**       | –   | 0.14864 | 0.15315 | 0.14438 | 0.17106 | **0.00375** | 3263 |
| **BPRMF + SI**  | –   | 0.16838 | 0.17112 | 0.16572 | 0.19500 | 0.00635 | 3552    |
| **WRMF**        | –   | 0.19514 | 0.19806 | 0.19231 | 0.22768 | 0.00454 | 766     |
| **WRMF + SI**   | –   | 0.19494 | 0.19782 | 0.19214 | 0.22773 | 0.00450 | 759     |
| **KG-AUTOENCODER** | 5   | 0.18857 | 0.18551 | 0.19173 | 0.21941 | 0.01835 | <u>5214</u> |
|                 | 10  | 0.21268 | 0.21009 | 0.21533 | 0.24945 | 0.01305 | 3350    |
|                 | 20  | 0.22886 | 0.22684 | 0.23092 | 0.27147 | 0.01015 | 2417    |
|                 | 40  | 0.23675 | 0.23534 | 0.23818 | 0.28363 | 0.00827 | 1800    |
|                 | 50  | 0.23827 | 0.23686 | 0.23970 | 0.28605 | 0.00780 | 1653    |
|                 | 100 | **0.23961** | **0.23832** | **0.24090** | **0.28924** | <u>0.00662</u> | 1310 |
| **AMAZON DIGITAL MUSIC** ||||||||
| **AUTOENCODER** | –   | 0.00060 | 0.00035 | 0.00200 | 0.00102 | 0.33867 | **3559** |
| **BPRMF**       | –   | 0.01010 | 0.00565 | 0.04765 | 0.02073 | **0.00346** | 539 |
| **BPRMF + SI**  | –   | 0.00738 | 0.00413 | 0.03480 | 0.01624 | 0.06414 | 2374    |
| **WRMF**        | –   | 0.02189 | 0.01236 | 0.09567 | 0.05511 | 0.01061 | 103     |
| **WRMF + SI**   | –   | 0.02151 | 0.01216 | 0.09325 | 0.05220 | 0.01168 | 111     |
| **KG-AUTOENCODER** | 5   | 0.01514 | 0.00862 | 0.06233 | 0.04365 | <u>0.03407</u> | 3378 |
|                 | 10  | 0.01920 | 0.01091 | 0.07994 | 0.05421 | 0.05353 | 3449    |
|                 | 20  | 0.02233 | 0.01267 | 0.09385 | 0.06296 | 0.08562 | 3523    |
|                 | 40  | 0.02572 | 0.01460 | 0.10805 | 0.06980 | 0.14514 | 3549    |
|                 | 50  | 0.02618 | 0.01486 | 0.10974 | 0.07032 | 0.17192 | <u>3549</u> |
|                 | 100 | **0.02835** | **0.01608** | **0.11964** | **0.07471** | 0.24859 | 3448 |
| **LIBRARYTHING** ||||||||
| **AUTOENCODER** | –   | **0.01562** | **0.01375** | **0.01808** | **0.01758** | 0.07628 | 2328 |
| **BPRMF**       | –   | 0.01036 | 0.00954 | 0.01134 | 0.01001 | 0.06764 | 3140    |
| **BPRMF + SI**  | –   | 0.01065 | 0.00994 | 0.01148 | 0.01041 | 0.10753 | **4946** |
| **WRMF**        | –   | 0.01142 | 0.01071 | 0.01223 | 0.01247 | **0.00864** | 439 |
| **WRMF + SI**   | –   | 0.01116 | 0.01030 | 0.01217 | 0.01258 | 0.00868 | 442     |
| **KG-AUTOENCODER** | 5   | 0.00840 | 0.00764 | 0.00931 | 0.00930 | 0.13836 | <u>4895</u> |
|                 | 10  | 0.01034 | 0.00930 | 0.01163 | 0.01139 | 0.07888 | 3558    |
|                 | 20  | 0.01152 | 0.01029 | 0.01310 | 0.01248 | 0.04586 | 2245    |
|                 | 40  | 0.01195 | 0.01073 | 0.01347 | 0.01339 | 0.02800 | 1498    |
|                 | 50  | 0.01229 | 0.01110 | 0.01378 | 0.01374 | 0.02403 | 1312    |
|                 | 100 | <u>0.01278</u> | <u>0.01136</u> | <u>0.01461</u> | <u>0.01503</u> | <u>0.01521</u> | 873 |

Table 3: Experimental Results